\documentclass{emulateapj} 
\usepackage{natbib}
\usepackage{latexsym,amssymb}
\usepackage{apjfonts}

\newcommand       \be           {\begin{equation}}
\newcommand       \ee           {\end{equation}}
\newcommand       \bea          {\begin{eqnarray}}
\newcommand       \eea          {\end{eqnarray}}

\newcommand       \kms		{\,{\rm km \,\, s}^{-1}}

\newcommand       \pc		{\,{\rm pc }}
\newcommand       \yr		{\,{\rm yr }}

\newcommand       \Myr		{\,{\rm Myr }}

\newcommand       \kpc		{\,{\rm kpc }}

\newcommand       \Rg		{R_{\rm GMC}}
\newcommand       \Mg		{M_{\rm GMC}}
\newcommand       \vg		{v_{\rm GMC}}
\newcommand       \Mcl		{M_{\rm cl}}
\newcommand       \mcl		{M_{\rm cl}}
\newcommand       \rcl		{r_{\rm cl}}

\newcommand       \Rb		{r_{\rm B}}
\newcommand       \rb		{r_{\rm B}}
\newcommand       \ms		{M_{*}}

\newcommand       \tauff	{\tau_{\rm ff}}
\newcommand       \tcl	        {\tau_{\rm cl}}
\newcommand       \taug	        {\tau_{\rm GMC}}
\newcommand       \eff		{\epsilon_{\rm ff}}
\newcommand       \etaG		{\eta_{\rm ff,G}}

\newcommand{\xclump}{R_{\rm cl}}
\newcommand{\rbondi}{\rb}
\newcommand{\rgmc}{\Rg}
\newcommand{\rhoconst}{\alpha_{\rho}}
\newcommand{\vturb}{v_{\rm T}}
\newcommand{\xclumphat}{\hat{R}_{\rm cl}}
\newcommand{\rbondihat}{\hat{R}_{\rm B}}
\newcommand{\Msun}{M_{\odot}}

\shorttitle{STAR FORMATION IN MASSIVE CLUSTERS}
\shortauthors{Murray \&  Chang}

\begin{document}

\title{STAR FORMATION IN MASSIVE CLUSTERS VIA BONDI ACCRETION} 

\author{Norman Murray\altaffilmark{1,2} \& Philip  Chang\altaffilmark{1,3}}
\altaffiltext{1}{Canadian Institute for Theoretical Astrophysics, 60
St.~George Street, University of Toronto, Toronto, ON M5S 3H8, Canada;murray@cita.utoronto.ca (NM); pchang@cita.utoronto.ca (PC)}
\altaffiltext{2}{Canada Research Chair in Astrophysics}
\altaffiltext{3}{Department of Physics, University of Wisconsin-Milwaukee, 1900 E. Kenwood Blvd., Milwaukee, WI 53211, USA}


\begin{abstract} 
Essentially all stars form in giant molecular clouds (GMCs). However, inside GMCs, most of the gas does not participate in star formation; rather, denser gas accumulates in clumps in the GMC, with the bulk of the stars in a given GMC forming in a few of the most massive clumps. In the Milky Way, these clumps have masses $\mcl\lesssim 5\times10^{-2}$ of the GMC, radii $\rcl\sim1\pc$, and free-fall times $\tcl\sim 2\times10^5\yr$. We show that clumps inside giant molecular clouds should accrete at a modified Bondi accretion rate, which depends on clump mass as $\dot M_{cl}\sim M_{cl}^{5/4}$. This rate is initially rather slow, usually slower than the initial star formation rate inside the clump (we adopt the common assumption that inside the clump, $\dot M_*=\eff M_{cl}/\tau_{cl}$, with $\eff\approx0.017$). However,
after $\sim2$ GMC free-fall times $\taug$, the clump accretion rate accelerates rapidly; formally, the clump can accrete the entire GMC in $\sim3\taug$. At the same time, the star formation rate accelerates, tracking the Bondi accretion rate. If the GMC is disrupted by  feedback from the largest clump, half the stars in that clump form in the final $\taug$ before the GMC is disrupted. The theory predicts that the distribution of effective star formation rates, measured per GMC free-fall time, is broad, ranging from $\sim0.001$ up to $0.1$ or larger and that the mass spectrum of star clusters is flatter than that of clumps, consistent with observations.
\end{abstract}



\section{INTRODUCTION}

The star formation rate (SFR) is a fundamental parameter of disk galaxies. It is well characterized on galactic disk scales by the Kennicutt-Schmidt relations \citep{1989ApJ...344..685K,1998ApJ...498..541K}. These relations come in two forms. The first form is that of a correlation between the surface density of star formation $\dot \Sigma_*$ (in solar masses per year or grams per second) and that of gas, $\Sigma_{\rm gas}$,
\be  \label{eqn: Kennicutt_mass}
\dot \Sigma_*= A \Sigma_{\rm gas}^\alpha,
\ee  
with $\alpha\approx1.4$. The second form  relates
$\dot\Sigma_*$ to the surface density of gas via the disk dynamical time $\Omega=v_c/R_d$,
\be  \label{eqn: kennicutt_time}
\dot\Sigma_* = \eta\Omega\Sigma_{\rm gas},
\ee  
where $R_d$ is the disk radius and $v_c$ is the circular velocity of the galaxy. The dimensionless parameter has an observationally determined value $\eta\approx0.017$. 
More recent work has refined these relations, particularly at low surface densities \citep{2008AJ....136.2782L,2008AJ....136.2846B}, but in galaxies where the bulk of the gas is molecular, Equations (\ref{eqn: Kennicutt_mass}) and (\ref{eqn: kennicutt_time}) remain valid.

There are numerous theoretical explanations of these relations, relying on very different  physics, including support by magnetic fields \citep{1976ApJ...207..141M,1983ApJ...273..202S}, suppression of collapse by supersonic turbulence \citep{1995MNRAS.277..377P,2005ApJ...630..250K}, energy feedback from stars and supernovae, and momentum feedback from massive stars \citep{2010ApJ...709..191M}. Which of these explanations is correct, if any, is currently still under debate.

Other work has focused on relating the rate of star formation to gas surface or volume density on sub-disk scales, ranging from $\sim1\kpc$ \citep{2010ApJ...722.1699S} in nearby galaxies, to $\sim100\pc$ in the Milky Way \citep{1988ApJ...334L..51M,1990ApJ...354..492M,1991ASPC...20...45E,2010ApJ...724..687L,2011ApJ...729..133M}. These studies find relations similar in form to Equation (\ref{eqn: kennicutt_time}). The mean value of the coefficient, averaged over many areas or star forming regions, is consistent with the global value $\eta\approx 0.017$. However, the dispersion of $\eta$ appears to vary with the scale on which the star formation is probed: measured values ranges from $\eta<10^{-3}$ to $\eta\approx0.5$ -- ranging almost four orders in magnitude.

It is well established that not all the gas in a galaxy participates in star formation; star formation takes place only in molecular gas, \citep[e.g.][]{2011arXiv1105.4605S}. Further, even inside a giant molecular cloud (GMC), not all the gas participates; rather, stars form primarily in high density gas, often in the form of ``clumps'' and filaments \citep[e.g.][]{2010A&A...518L.100M}. Another way to see this is that star formation rates scale non-linearly with CO luminosity, which traces rather low density gas, while they scale linearly with HCN, which traces high density gas \citep{2004ApJ...606..271G}. In local spirals like the Milky Way, the fraction of HCN gas relative to molecular (CO emitting) gas is roughly several to ten percent \citep[e.g][]{2010ApJS..188..313W}.

In this paper we present a simple theory for the rate of star formation in GMCs, in which gravity plays the dominant role. We assume that the gas in a GMC is turbulently supported, i.e., the GMC is in rough virial equilibrium. The turbulence seeds the cloud with parsec scale clumps having masses $\delta\equiv\mcl/\Mg\approx10^{-3}$; \citet{2010ApJS..188..313W} refer to these objects as ``massive dense clumps''. These clumps then grow by accretion. While feedback from one or two of the most massive clumps eventually overcomes gravity and disrupts the GMC hosting the star formation \citep[see for instance,][]{2010ApJ...709..191M,2011ApJ...729..133M}, we ignore the effects of feedback. We argue that the feedback affects the accretion only in the last stages of clump growth; both the clump accretion and star formation rate accelerate rapidly, and hence so do the effects of feedback.

By analogy with Equation (\ref{eqn: kennicutt_time}), we define the effective GMC star formation efficiency per free-fall time as
\be \label{eqn: epsilon_effective} 
\etaG \equiv \taug {\dot M_*\over \Mg},
\ee 
where $\taug$ is the free-fall time of the GMC. Our gravity-dominated theory of star formation in GMCs predicts that $\etaG$ is a strong function of time; it is small immediately after the GMC is assembled and for a substantial fraction of a GMC free-fall time, but then increases rapidly; we show that this explains the large dispersion of $\eta$ on small scales referred to above.

This paper is organized as follows. In \S \ref{sec: bondi} we describe accretion onto massive clumps in GMCs; the mass accretion rate accelerates as a clump grows. We show that the star formation rate in the clump tends to track the mass accretion rate. It follows that $\etaG$ is time dependent, even if the star formation rate per free fall time ($\eff$, defined in Equation (\ref{eqn:turbulence}) below) is constant for the parsec scale clumps of molecular gas which form individual star clusters. In \S \ref{sec:compare} we compare our results to observations of star formation in the Milky Way and nearby galaxies; the theory predicts a large spread in the apparent star formation rate in GMCs, and a flatter mass distribution of star clusters compared to star forming clumps. In \S \ref{sec: discussion} we compare our results with recent numerical simulations, and we briefly discuss the effects of vorticity, arguing that they do not limit the rate of accretion on to dense clumps. Finally, we close with our conclusions.

\section{BONDI ACCRETION IN GIANT MOLECULAR CLOUDS}\label{sec: bondi}
Simulations of supersonic turbulence suggest that it prevents gas from rapidly fragmenting and turning into stars on the local dynamical time
\citep{1995MNRAS.277..377P,2000ApJ...535..887K,2004ApJ...605..800L}. Based on these and
similar results, numerous authors have suggested that, over the local dynamical or free fall time, only a small fraction of the gas in a given region is turned into stars
\citep{1995MNRAS.277..377P,2005ApJ...630..250K}. In particular, if the mean density in a region is $\bar\rho$, the local free-fall time is 
\be  
\tauff\equiv\sqrt{3\pi\over 32 G\bar\rho}.
\ee  
Then the star formation rate is given by
\be  \label{eqn:turbulence} 
{dM_*\over dt}=\eff{M_g\over \tauff},
\ee  
with $\eff\approx0.017$.  While we are agnostic about the applicability of this prescription, we will adopt it here.

We now show that the local star formation rate given by Equation (\ref{eqn:turbulence}) does not limit the overall rate of star formation in GMCs very significantly. This results from two facts; first, for dense enough clumps, the star formation timescale is shorter than $\eff$ times the free-fall time of the GMC, and second, star forming clumps control the dynamics of the gas in their vicinity through their gravity, as we now show.


GMCs are observed to harbor both massive gas clumps and massive star clusters. For example, in the Milky Way, the most massive GMCs have $\Mg\approx3\times10^6M_\odot$ and $\Rg\approx100\pc$, e.g., \citet{1988ApJ...324..248B,1988ApJ...331..181G}. These massive GMCs contain dense clumps with $\Mcl\approx 10^4M_\odot$ \citep{2001ApJ...551..747S,2010ApJS..188..313W} and star clusters with similar or larger masses \citep{2010ApJ...709..424M}. We show that the masses of these gas clumps initially grow slowly, but the growth rate then accelerates rapidly, so that, left unchecked, the largest clump would consume the GMC in a few $\tauff$.

Consider a giant molecular cloud of radius $\Rg$ and mass $\Mg$, which we will assume is near virial equilibrium. The velocity on the scale $\Rg$ is $\vg=\sqrt{G\Mg/\Rg}$, while the turbulent velocity inside the cloud is given by
\be  
v_T(r)=\vg\left({r\over\Rg}\right)^p
\ee  
where $p\approx1/2$, by one of Larson's laws \citep{1981MNRAS.194..809L,1987ApJ...319..730S}.  In the most massive Milky Way GMCs, $v_T(\Rg)=\vg\approx6\kms$; in more rapidly star forming galaxies such as 
the $z=2$ galaxy BX 482, $\vg\approx 50\kms$.

Now consider a clump embedded in the GMC with a total mass $\mcl$ (including a stellar mass $\ms$) and radius $\rcl$. The gravity of the cluster controls the flow of the surrounding gas out to the Bondi radius \citep{1952MNRAS.112..195B}
\be  
\Rb={G\mcl\over v_T^2(\Rb)}.
\ee  
Solving for $\Rb$,
\be  
\Rb=\left({\mcl\over \Mg}\right)^{1/2}\Rg=\mu^{1/2}\Rg,
\ee  
where we have defined $\mu\equiv\mcl/\Mg$. The turbulent velocity at the Bondi radius is
\be 
v_T(\Rb)=\mu^{1/4}\vg.
\ee 
The associated Bondi accretion rate is
\be \label{eqn:mdot Bondi} 
\dot M_{\rm B}=4\pi\Rb^2v_T(\Rb)\rho.
\ee  

The density $\rho$ is a function of position inside the GMC. Assuming spherical symmetry, the literature contains suggested functions of the form
\be  \label{eq:density}
\rho(r)=C\bar\rho\left({\Rg\over r}\right)^\beta,
\ee  
where $C=(3-\beta)/3$, and $\bar\rho\equiv 3\Mg/(4\pi\Rg^3)$ is the mean density of the GMC.  We will consider the case of $\beta = 0$ (constant density), $1$, and $2$ (isothermal distribution).  

It is likely that clumps form where gas flows shock and subsequently cool. While such shocks can occur anywhere in the interior of the GMC, there is some observational evidence that star clusters may be biased toward the centers of GMCs, e.g., HII regions are centrally concentrated in their host GMCs \citep{1987ApJS...63..821S}. We will simply assume that our clump forms at a distance $R_{cl}$ from the center of the GMC.  

\subsection{Constant density GMCs}
For a constant density cloud ($\beta = 0$), the position of the clump does not affect the accretion rate (so long as the Bondi radius of the clump does not exceed the distance to the surface of the GMC) so we will focus on this case first. The accretion rate is then
\be  \label{eqn:mdot Bondi const} 
\dot M_{\rm Bondi}={3\pi\over 2\sqrt{2}}\mu^{5/4}{\Mg\over \taug},
\ee  
where $\taug$ is the free-fall time of the GMC.

Milky Way GMCs often have many such clumps over a ranges of masses \citep[see for instance][]{2001ApJ...551..747S}. We focus on the largest clump because the $\mcl^{5/4}$ dependence on the growth rate in Equation (\ref{eqn:mdot Bondi const}) implies that the largest clump also grows the fastest. We assume that clumps are born with an initial mass $M_0=\delta\eff \Mg$, and that the sum of all the gravitationally bound clumps produced over a free-fall time is $\sim\eff\Mg$, i.e., that the sum over clumps $\Sigma \delta=1$. 

The time evolution of the mass of an individual accreting clump is given by
\be  \label{eqn:final} 
\mcl(t)=M_0
\left[
1-{3\pi\over8\sqrt{2}}{t\over \taug}
\left({M_0\over \Mg}\right)^{1/4}
\right]^{-4}.
\ee  
The denominator in Equation (\ref{eqn:final}) vanishes at a finite time; before that happens, the GMC is consumed, after a time
\bea  
t_{\rm final}&=&{8\sqrt{2}\over3\pi}
\left[
\left({\Mg\over M_0}\right)^{1/4} - 1
\right]
\taug\\
&\approx&3.5\left({\delta\over 0.2}\right)^{-1/4}
\left({\eff\over 0.02}\right)^{-1/4}\taug.
\eea  

We interpret $t_{\rm final}$ as a hard upper limit on the lifetimes of GMCs; it is the time it takes the clump to completely accrete the GMC.  While this calculation ignores both angular momentum considerations and feedback processes, we argue below that the GMC will be disrupted by stellar feedback before $t_{\rm final}$ is reached and that angular momentum does not play a significant role.

\subsubsection{Star Formation in Bondi Clumps}\label{sec:star formation}

The discussion in the previous section follows the evolution of a clump as it accretes gas from its parent GMC over several $\taug$. Now we refine our picture, and discuss the evolution of stellar mass in the clump. The clump mass is divided into stars and gas; we denote the stellar mass as $M_*$, and we define the gas fraction 
\be \label{eqn:fgas}
f_g\equiv{\mcl-M_*\over \mcl}.
\ee 
We use the star formation law from Equation (\ref{eqn:turbulence}),
\be   
{dM_*\over dt}=\eff{f_g\mcl\over\tcl},
\ee   
where $\tcl$ is the free-fall time of the cluster. The free-fall time of the cluster is given by 
\begin{equation}
\tcl = \sqrt{\frac{3\pi}{32G\rho_{\rm cl}}} = \taug \sqrt{\frac {(R_{\rm cl}/R_{\rm GMC})^3} {\mu}}.
\end{equation}

It is useful to re-scale all the cluster masses (total, stellar, and gas) by the GMC mass, so in addition to $\mu = \mcl/\Mg$, we have $\mu_g = f_g\mu$ and $\mu_* = M_*/M_{\rm GMC}$.  The evolution equations of the gas and the stars are then given by
\begin{eqnarray}
\dot{\mu}_* &=& \eff {\mu_g \over \tcl}, \label{eq:stars}\\
\dot{\mu}_g &=& \dot{\mu} - \eff {\mu_g \over \tcl}. \label{eq:gas}
\end{eqnarray}

Combining Equations (\ref{eqn:fgas}) and (\ref{eq:gas}) we have
%
%
\be \label{eq: gas fraction} 
{df_g\over dt} = (1-f_g){\dot\mu\over\mu}-\eff{f_g\over\tcl}.
\ee  

We can look for a fixed point of the last equation by setting the left-hand side of this equation to zero:
\be \label{eq: fixed point}
f_{g,{\rm fixed}} = {1\over 1+g(\mu)},
\ee 
where
\be \label{eq: g mu}
g(\mu)\equiv {2\sqrt{2}\over 3\pi}\eff
\left({\Rg\over \rcl(\mu)}\right)^{3/2}\mu^{1/4}.
\ee 

We have called $f_{g,{\rm fixed}}$ a fixed point, although it is only fixed if $\mu$ is a constant. In fact, $\dot\mu/\mu\sim \mu^{1/4}$ is an increasing function of time, so Equation (\ref{eq: fixed point}) does not give the instantaneous value of $f_g$. As the cluster grows $\mu$ will increase, resulting in a change in $f_{g*}$. The actual value of $f_g$ will be slightly different than $f_{g*}$, with the magnitude of the off-set controlled by the ratio of the star formation time (from Equation (\ref{eqn:turbulence})) to the clump accretion time (from Equation (\ref{eqn:mdot Bondi})). 

It is easy to show that this psuedo-fixed point is an attractor, in the sense that if $f_g$ is initially smaller than $f_{g*}$ the gas fraction will increase, approaching $f_{g*}$ asymptotically, while $f_g$ will decrease if it is larger than $f_{g*}$.

\subsubsection{The mass radius relation for clumps}
In order to integrate Equations (\ref{eqn:mdot Bondi}) or (\ref{eqn:mdot Bondi const}),  (\ref{eq:stars}), and (\ref{eq:gas}), we need to know $\tcl$, which requires knowledge of the mass-radius relation for massive clumps in GMCs. We have not been able to find much information in the literature, so as a proxy we use the mass-radius relation for young star clusters, under the assumption that the radius of such clusters reflects, to some extent, the radius of the parent gas clumps.

The half light radii of star clusters \citep{2005ApJ...618..237W,2009ApJ...691..946M,2009gcgg.book..103S} and the effective radii of clumps \citep{2001ApJ...551..747S} have characteristic values of $1-3\pc$ for masses below $10^6M_\odot$ so we adopt a fiducial value of $1\pc$. More massive star clusters have radii given by \citep[see][]{2009ApJ...691..946M}
\be   
\rcl(\mcl)\approx 1\left({\mcl\over 10^6M_\odot}\right)^{3/5}\pc.
\ee   
These relations are clearly very rough approximations; better observational data on the mass-radius relation for massive clumps in GMCs would be of great value.


Taking parameters from observed massive star clusters, we find that $g(\mu)$ ranges from a minimum value of $g(\mu)\approx0.1$ for low mass ($10^3M_\odot$) clusters in $30-100\pc$, $10^5-10^6M_\odot$ GMCs in local galaxies up to a maximum value of $g\approx 50$, for $10^6M_\odot$ clusters in the $\Rg=1\kpc$, $10^8M_\odot$ GMCs in high redshift star forming galaxies. The corresponding limits for the gas fraction are $0.02\lesssim f_{g,{\rm fixed}}\lesssim0.9$.

\subsubsection{Low mass star clusters in local galaxies}

\begin{figure}
\begin{center}
\includegraphics[width=.5\textwidth]{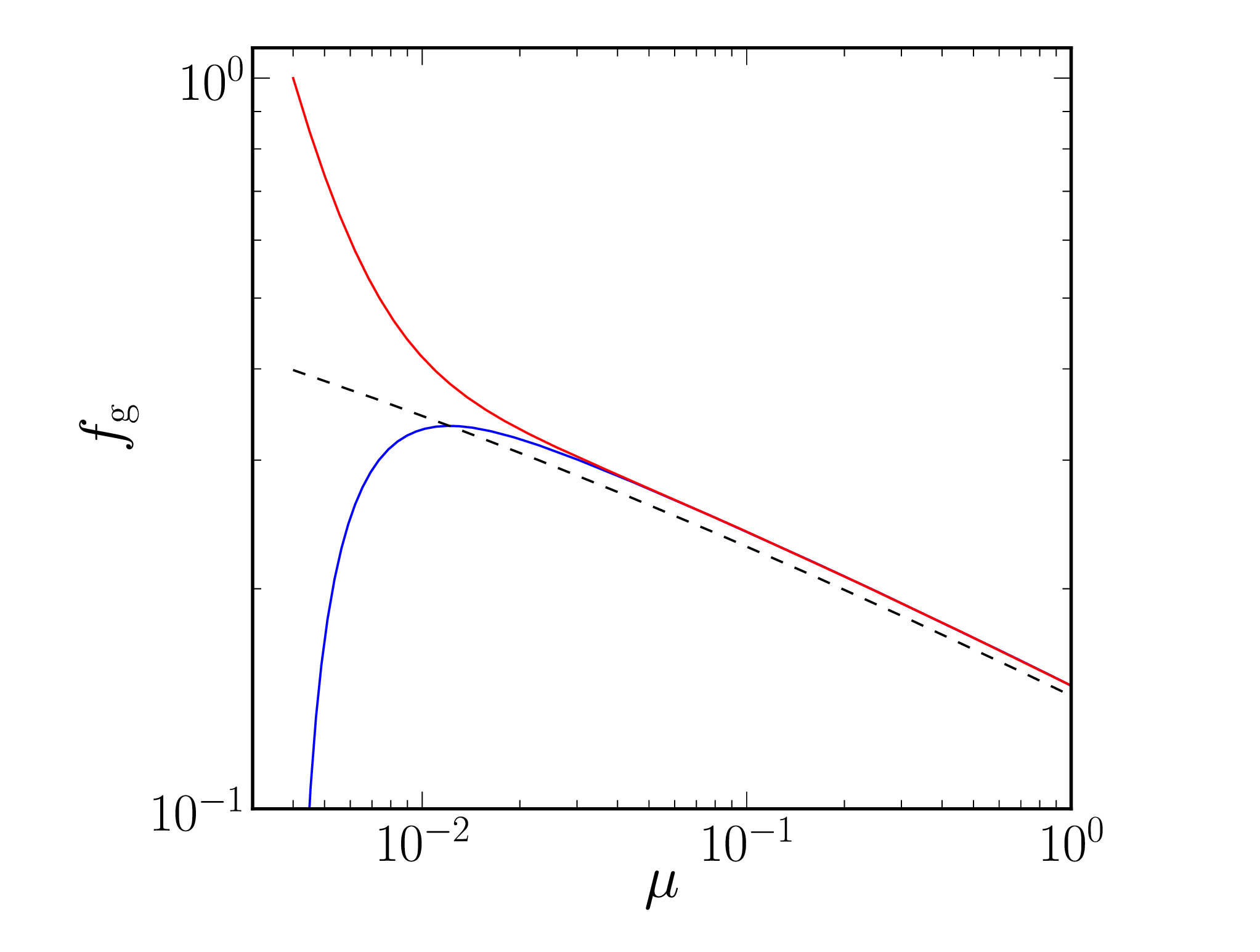}
\end{center}
\caption{The gas fraction of an accreting clump plotted as a function of the ratio of clump mass to GMC mass, $\mu$. The upper solid line shows the result of starting with a pure gas clump, while the lower solid line shows the result of starting with a purely stellar clump. The two lines rapidly converge as the clump mass grows. The dashed line shows the approximate fixed point gas fraction $f_{g*}$ given by Equations (\ref{eq: fixed point}) and (\ref{eq: g mu}), which applies in the limit of a very slowly growing clump $\dot\mu<<1$.
}
\label{fig: gas fraction r fixed} 
\end{figure}

The result of integrating Equation (\ref{eq: gas fraction}) for a cluster with $\mcl<10^6M_\odot$ in a $10^6M_\odot$, $\Rg=100\pc$ GMC is illustrated in Figure \ref{fig: gas fraction r fixed}. This shows the result of two numerical integrations, one with an initial gas fraction of one, and a second with an initial gas fraction of zero. The dashed line shows $f_{g,{\rm fixed}}(\mu)$; the fact that both numerical integrations converge to a value slightly above $f_{g,{\rm fixed}}$ shows both that Equation (\ref{eq: gas fraction}) has an attractor, and that $f_{g,{\rm fixed}}$ is a good approximation for that attractor. Since $\rcl$ is taken to be fixed, $g(\mu)$ is monotonically increasing, but it never exceeds unity, so $f_g(\mu)$ is monotonically decreasing and rather featureless.

Note that $f_g$ varies slowly with $\mu$, only by a factor of 2 (after the initial transient), while $\mu$ varies by nearly two orders of magnitude. This shows that the star formation rate traces the clump mass accretion rate.  In other words, the star formation rate is set by the clump accretion rate rather than by the properties of the local turbulence.

\subsubsection{Massive star clusters}
For $\mcl>10^6M_\odot$, the clump radius grows with increasing mass. In this case, $g(\mu)$ decreases with increasing $\mu$, so that $f_g$ increases. Hence, the star formation rate lags behind the Bondi accretion rate. However, $1-f_g\approx g(\mu)$ is never very small, as noted above, so the star cluster mass is always a substantial fraction of the clump mass -- both increase rapidly as $t\to t_{\rm final}$. Once again, the star formation rate is set by the clump mass accretion rate rather than by turbulence.

\begin{figure}
\begin{center}
\includegraphics[width=.5\textwidth]{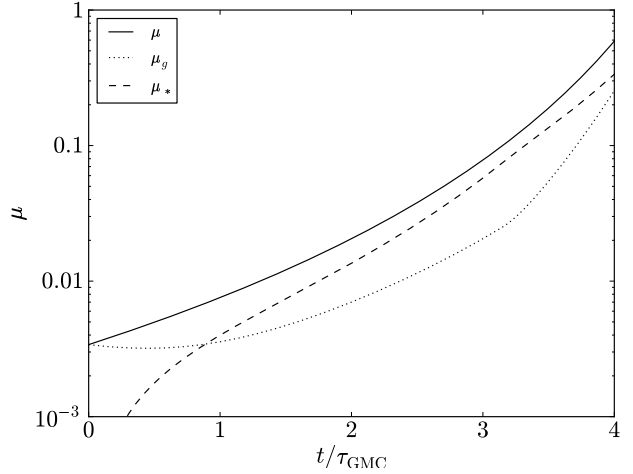}
\end{center}
\caption{
The total (solid line), stellar (dashed line), and gas (dotted line) mass fraction of a clump undergoing Bondi accretion, plotted as a function of $t/\taug$. The mass and radius of the host GMC are $\Mg=10^7M_\odot$ and $\Rg=100\pc$. The dynamic time of the GMC is $\taug = 4.5 $Myrs.  We ignore the effects of stellar feedback. The clump experiences a long period of quiescience, which is followed by a period of rapid growth.  The stellar mass initially tracks the clump mass very closely, but for $M_* > 10^6 \Msun$ (at $t/\taug\approx3.3$), the star formation rate grows less quickly than the accretion rate, so the stellar mass stops tracking the clump mass so closely. This is most easily seen by the change in slope of $\mu_g(t)$.  This change in behavior is due to the change in the clump mass-radius relation, which we have assumed is constant for star clusters with $M_*< 10^6\Msun$, but has radius increasing with mass for $M_*>10^6\Msun$.}
\label{fig:beta0} 
\end{figure}

In Figure \ref{fig:beta0}, we plot the evolution of a Bondi clump accreting from a constant density GMC, where $\Mg = 10^7\Msun$ and $\Rg = 100$ pc. We show the total (solid line), stellar (dashed line) and gas (dotted line) mass fractions for the Bondi clump. For most of the lifetime of the clump, it remains small.  However it grows rapidly at late times, consuming the GMC after $\sim4 \taug\approx 19 \Myr$, in line with our expectations for $t_{\rm final}$.  We also track the stellar and gas mass separately and note the difference in behavior for $\mu_* < 0.1$ ($M_*<10^6\Msun$) and $\mu_*>0.1$ ($M_* > 10^6\Msun$).  Initially the stellar mass tracks the clump mass very closely. Beyond $M_*>10^6\Msun$ the stellar mass still tracks the clump mass, but not as closely as in the fixed $\rcl$ regime.  This is especially evident in the kink in $\mu_g$.  


\subsection{Clump growth by Bondi Accretion: General Case}\label{sec:off-center}

We now consider the case of a clump located away from the center of a non-uniform GMC.  Suppose the clump is a distance $\xclump$ away from the center.\footnote{Note the difference between $\xclump$ and $\rcl$. $\xclump$ defines the radial position of the clump in the GMC, while $\rcl$ defines the size of the clump.} If $\xclump \gg \rbondi$, then our previous discussion holds -- the clump accretes from a (locally) nearly uniform medium.  If we relax this assumption, the density depends on the position at the surface of the Bondi sphere.  Recall that we assume a density profile of the GMC given by  $\rho(r) = \rho_0\left(r/\rgmc\right)^{-\beta}$.

We choose a second coordinate system centered on the clump, with the center of the GMC located on the z-axis a distance $\xclump$ away. A point on the surface of the Bondi sphere, described by the second coordinate system $(\rbondi, \theta, \phi)$ 
lies a distance $r(\rbondi, \theta, \phi)$ from the center of the GMC, where
\begin{equation}
r(\rbondi,\theta,\phi) = \sqrt{\rbondi^2 + \xclump^2 + 2\rbondi\xclump\cos\theta}.
\end{equation}

We assume that the turbulent velocity is homogeneous and isotropic, unlike the density. Using this assumption, the mass accretion rate is:
\begin{eqnarray}
\dot{M} &=& \int \rbondi^2 d\cos\theta d\phi \rho(r(\rbondi,\theta,\phi)) \rbondi^2 \vturb(\rbondi) \nonumber\\
&=& \frac{(3-\beta)\Omega_{\rm GMC}}{2} M_{\rm GMC} \mu^{5/4}\nonumber \\
&\times& \int_{-1}^1 d\cos\theta \left(\xclumphat^2 + \rbondihat^2 + 2\xclumphat\rbondihat\cos\theta\right)^{-\beta/2},
\end{eqnarray}
where $\rbondihat = \rbondi/R_{\rm GMC} = \sqrt{\mu}$, $\Omega_{\rm GMC} = \taug^{-1}$, and $\xclumphat = \xclump/R_{\rm GMC}$.  Performing the integral over $\cos\theta$, we find
\begin{eqnarray}
\dot{\mu} &=& \frac{(3-\beta)\Omega_{\rm GMC}}{2(2 - \beta)\xclumphat} \mu^{3/4}  \times \nonumber\\
&&\left[\left(\xclumphat^2 + \mu + 2\xclumphat\sqrt{\mu}\right)^{1-\beta/2} - 
\left(\xclumphat^2 + \mu - 2\xclumphat\sqrt{\mu}\right)^{1-\beta/2}\right], \label{eq:off-center-beta} 
\end{eqnarray}
for $\beta \neq 2$ and 
\begin{eqnarray}
\dot{\mu} = \frac{(3-\beta)\Omega_{\rm GMC}}{2\xclumphat} \mu^{3/4}\log\left(\frac {\xclumphat^2 + \mu + 2\xclumphat\sqrt{\mu}}{\xclumphat^2 + \mu - 2\xclumphat\sqrt{\mu}}\right),\label{eq:off-center-beta2} 
\end{eqnarray}
for $\beta = 2$.  

Equations (\ref{eq:off-center-beta}) and (\ref{eq:off-center-beta2}) define the evolution of an accreting clump at a position $\xclump$.  We note that Equation (\ref{eq:off-center-beta}) reduces to 
\begin{equation} \label{eq: general mdot}
\dot{\mu} = 3\rhoconst\Omega_{\rm GMC}\mu^{5/4}\xclumphat^{-\beta} \left[ 1 + O(\mu^{3/2})\right]
\end{equation}
for $\rb << \xclump$, which we recognize as Equation (\ref{eqn:mdot Bondi const}), where the background density is set by the position of the clump.  A similar limiting form is found from Equation (\ref{eq:off-center-beta2}).

\begin{figure*}
\begin{center}
\includegraphics[width=.48\textwidth]{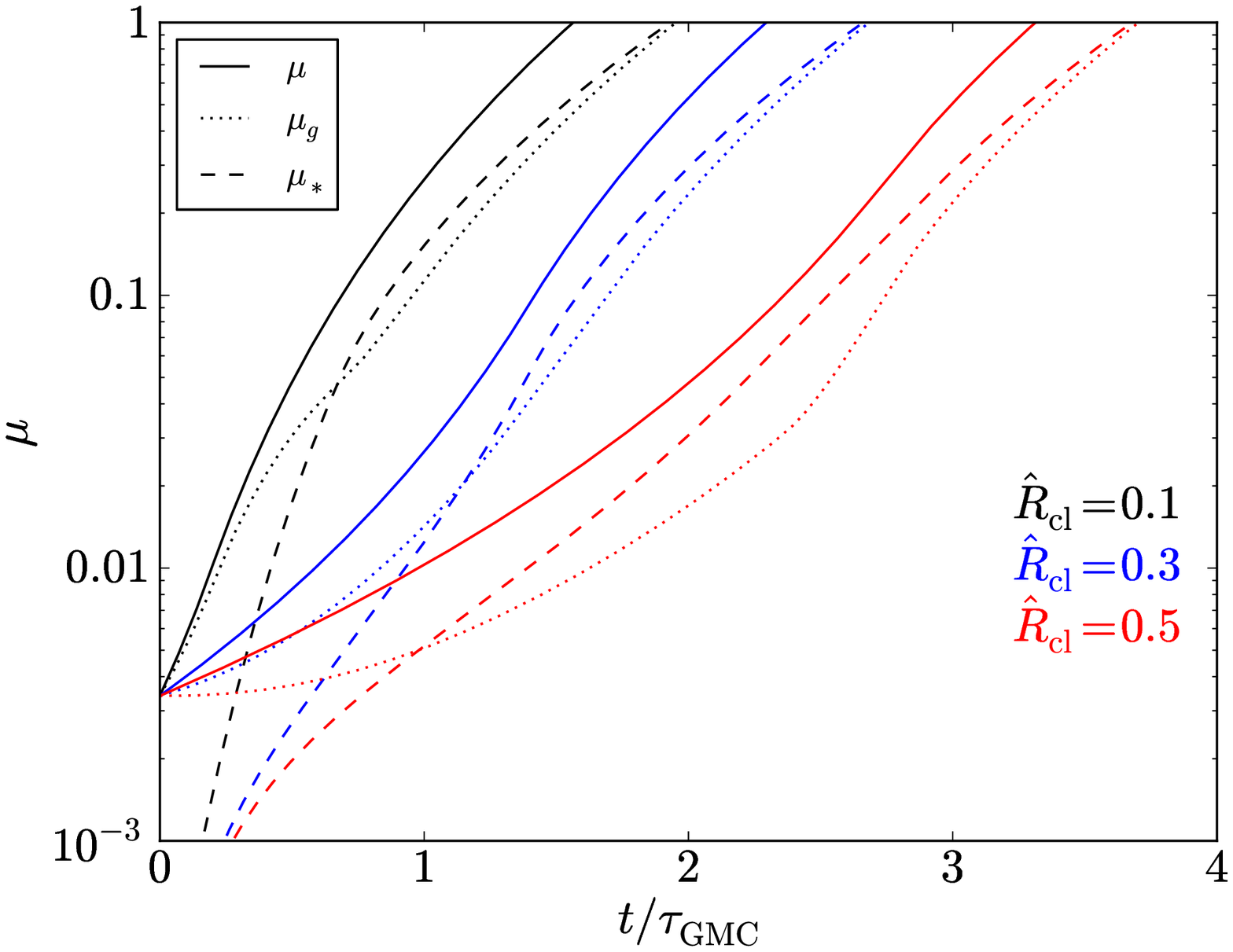}
\includegraphics[width=.48\textwidth]{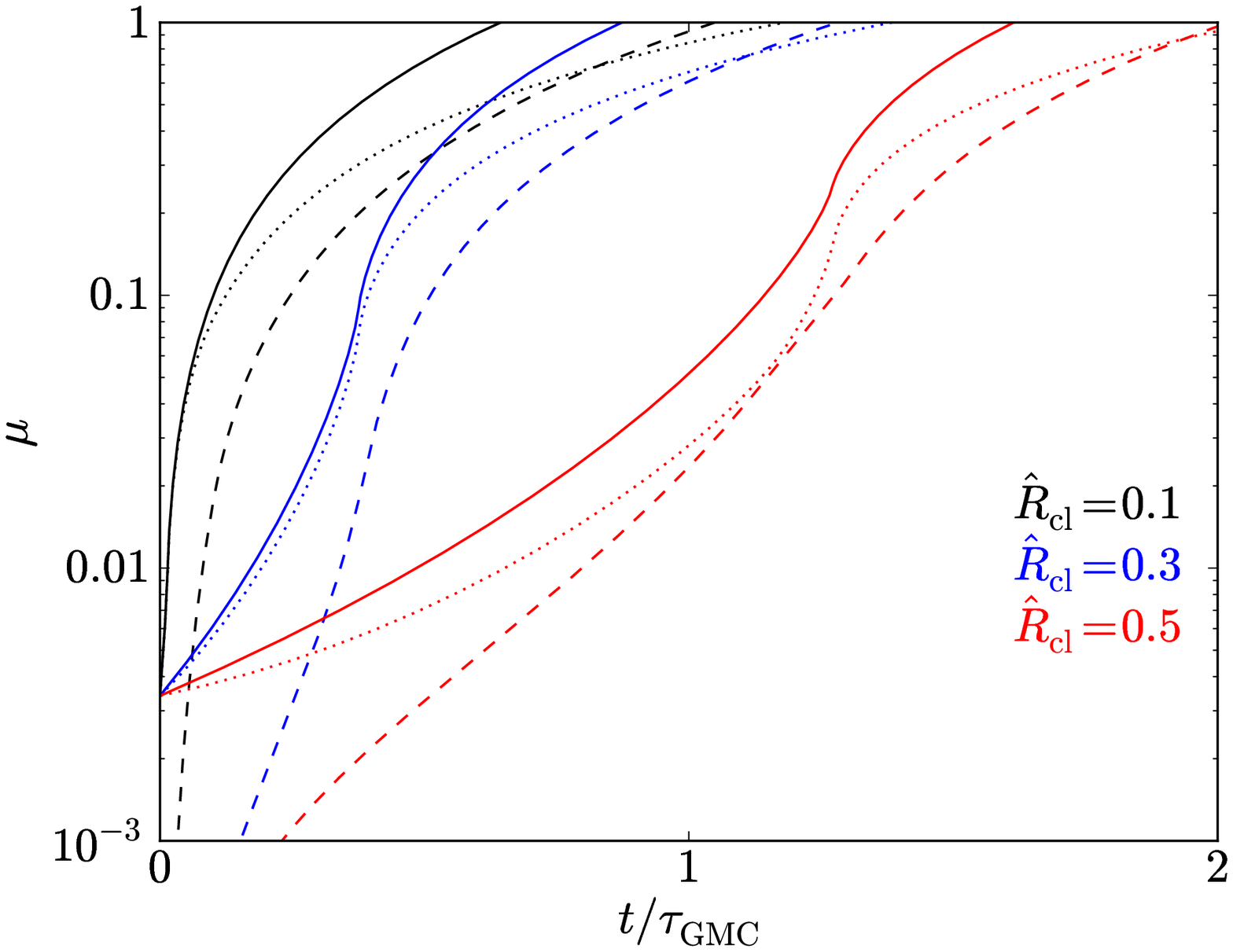}
\end{center}
\caption{Mass fraction, stellar fraction, and gas fraction of the clump as a function of time for $\beta = 1$ (left), and $2$ (right) for a GMC with $\Mg = 10^6\Msun$. We show $\xclumphat = 0.1$ (black lines), 0.3 (blue lines), and 0.5 (red lines). For a clump growing near the center ($r=0.1$), note that the clump consumes the entire GMC is $\approx 1.5 \taug$ for $\beta = 1$ and $0.5 \taug$ for $\beta = 2$.  Clumps that are off center grow slower, but even slowest growing clump consumes the entire GMC in $3.5 (1.5)\taug$ for $\beta = 1$ ($\beta=2$).  For off-center clumps ($r=0.3$ or larger), which, as noted below  Equation (\ref{eq:off-center-beta2}) reduces to the $\beta = 0$ case, the clump experiences a long period of quiescience, followed by a period of rapid growth, but the period over which growth occurs is much shorter than in the $\beta=0$ case.  In these cases, we have not included a $\xclumphat=0$ curve because it is nearly traced out by the $\xclumphat=0.1$ curve.
}
\label{fig:beta 1 and 2} 
\end{figure*}

It is straightforward to integrate Equations (\ref{eq:off-center-beta}) and (\ref{eq:off-center-beta2}), along with Equations (\ref{eq:stars}) and (\ref{eq:gas}).  The results are shown in Figures \ref{fig:beta0} (for $\rho(r)=const.$, i.e., $\beta = 0$) and \ref{fig:beta 1 and 2} for $\beta=1$ (left hand panel) and $\beta=2$ (right hand panel). Here, we show the total (solid line), stellar (dashed line) and gas (dotted line) mass fractions for the various Bondi clumps. For $\beta\ne0$ we present integrations at three example radii, $\xclumphat = $ 0.1 (black lines), 0.3 (blue lines), and 0.5 (red lines), to demonstrate that the clump mass accretion rate depends rather sensitively on the position of the clump in the GMC.\footnote{In Figure \ref{fig:beta 1 and 2}, we have not included a $\xclumphat=0$ curve because it is nearly traced out by the $\xclumphat=0.1$ curve.}

Clumps in centrally concentrated GMCs exhibit rapid growth. This is shown by the $\xclumphat=0.1$ curves (solid lines) of Figure \ref{fig:beta 1 and 2}.  As we have already discussed, the high densities of these central regions lead to rapid growth,  in which the entire GMC is accreted in a single (mean) GMC free-fall time $\taug$.  

Clumps that are significantly off-center, i.e., $\xclumphat=0.3$ and $0.5$, behave more like the constant density GMC case.  There is a period of slow growth, followed by a period of rapid growth. However, because these host GMCs  are more centrally concentrated than a constant density GMC, the time it takes an off-center clump to accrete a substantial fraction of the host GMC mass is less than that seen in the  constant density GMC by a factor of a few, when measured in units $\taug$.

The division between ``small'' and ``large'' $\xclumphat$ depends on the (initial) Bondi radius of the clump; since we are using $\mu=0.003$, the initial Bondi radius is $0.06$, so $\xclumphat=0.1$ is at the transition between rapidly accreting clumps and more slowly accreting clumps.

\section{Comparison to Observations}\label{sec:compare}

\subsection{Bondi Clumps and the observed rate of star formation in GMCs}\label{sec:sfr}

These results suggest a robust story for star formation in GMCs, independent of the host GMC density profile. First, much of the star formation happens in the most massive (or the few most massive) cluster(s). Second, most of the star formation occurs near the end of the GMC's lifetime. We expect that this rapid burst of star formation is stopped by the disruption of the host GMC.

Finally, the star formation rate is very different in GMCs than one would expect in models of star formation which essentially extend the Kennicutt relation (Equation (\ref{eqn: kennicutt_time})), valid on galactic scales, to the scale of individual GMCs, and assuming that the star formation rate follows a Poisson process over the entire cloud given by Equation (\ref{eqn:turbulence}) with random values of $\eta$ that average to $\eta\approx0.02$, and where the cluster mass is determined by the random confluence of turbulent statistics.  

Rather, in the scenario proposed here, turbulence generates clumps which subsequently grow by accretion. These clumps then experience runaway growth until stellar feedback ultimately unbinds the GMC, arresting the growth of all the clumps. This story highlights the importance of feedback on capping the ultimate star formation efficiency of roughly $10\%$ in the Milky Way. 

To make observational contact, we calculate distribution of observed star formation rates in GMCs in our model. The star formation rate in a given GMC is driven primarily by the age of that GMC. Given a collection of GMCs with randomly distributed ages, the observed SFR distribution is simply the star formation rate as a function of GMC age convolved with the distribution of GMC ages. Very roughly, the likelihood of finding a GMC with a star formation rate in a given range is proportional to the length of time the most massive clump spends in the relevant mass range. 

\begin{figure*}
\begin{center}
\includegraphics[width=.33\textwidth]{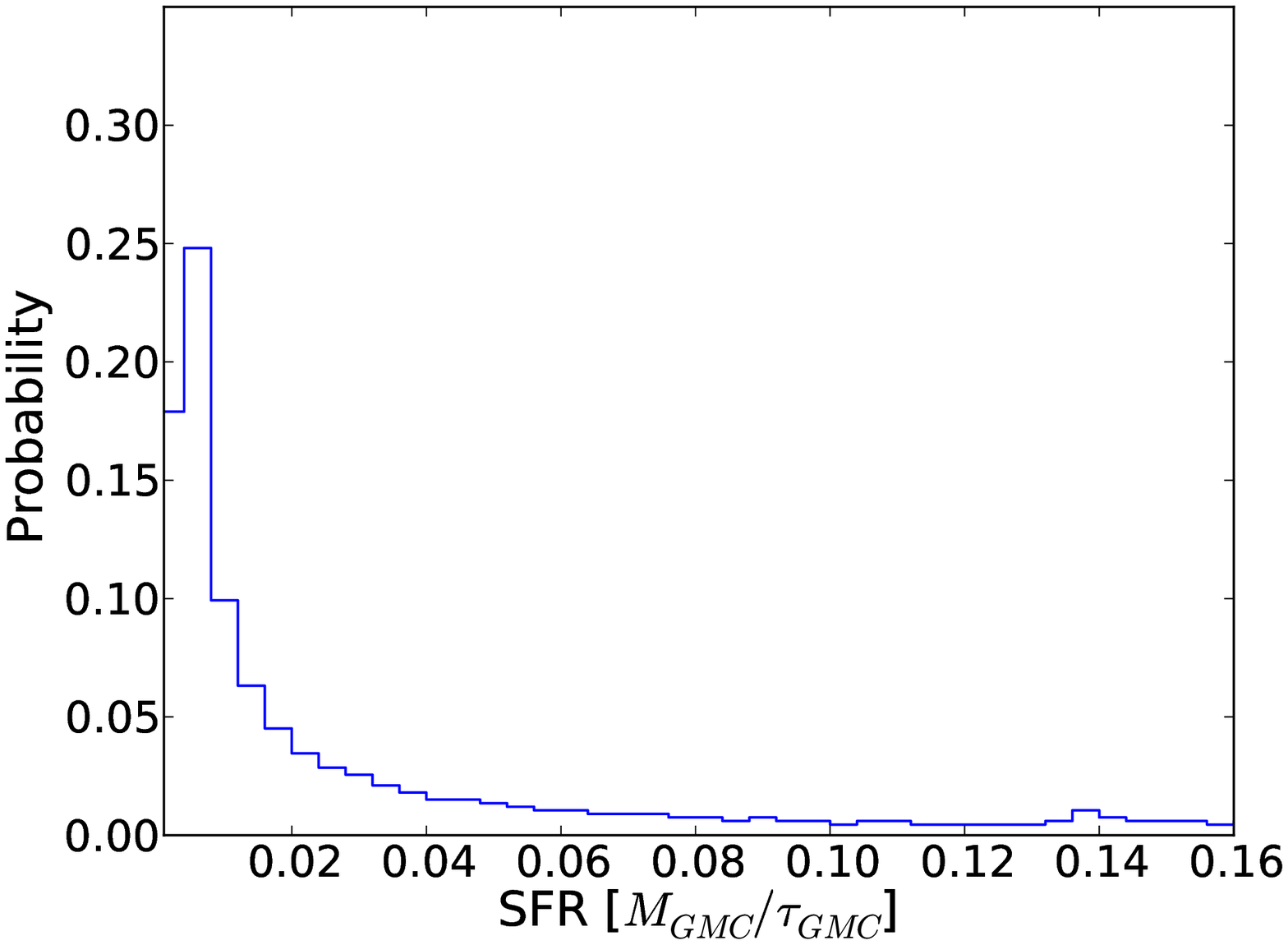}
\includegraphics[width=.33\textwidth]{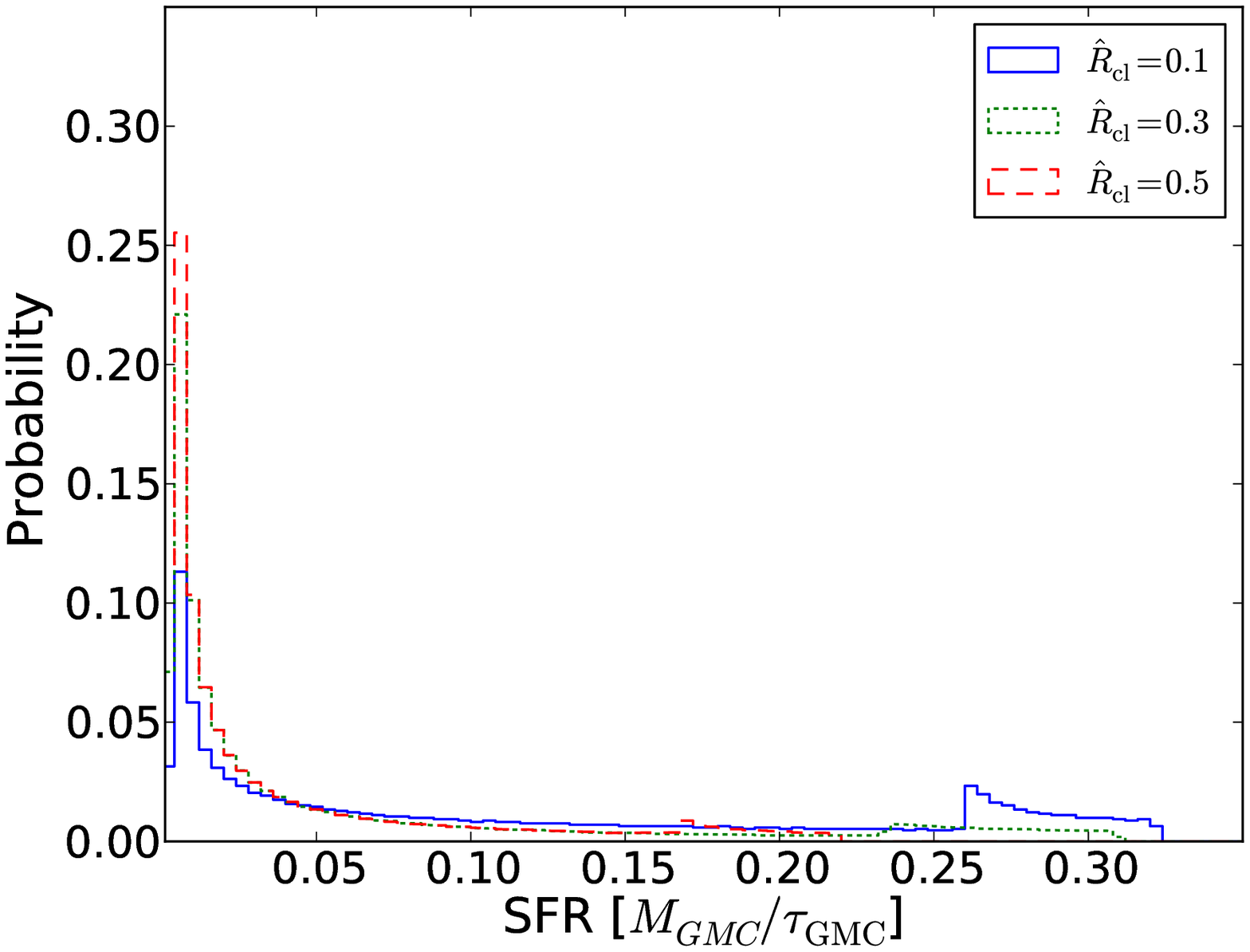}
\includegraphics[width=.33\textwidth]{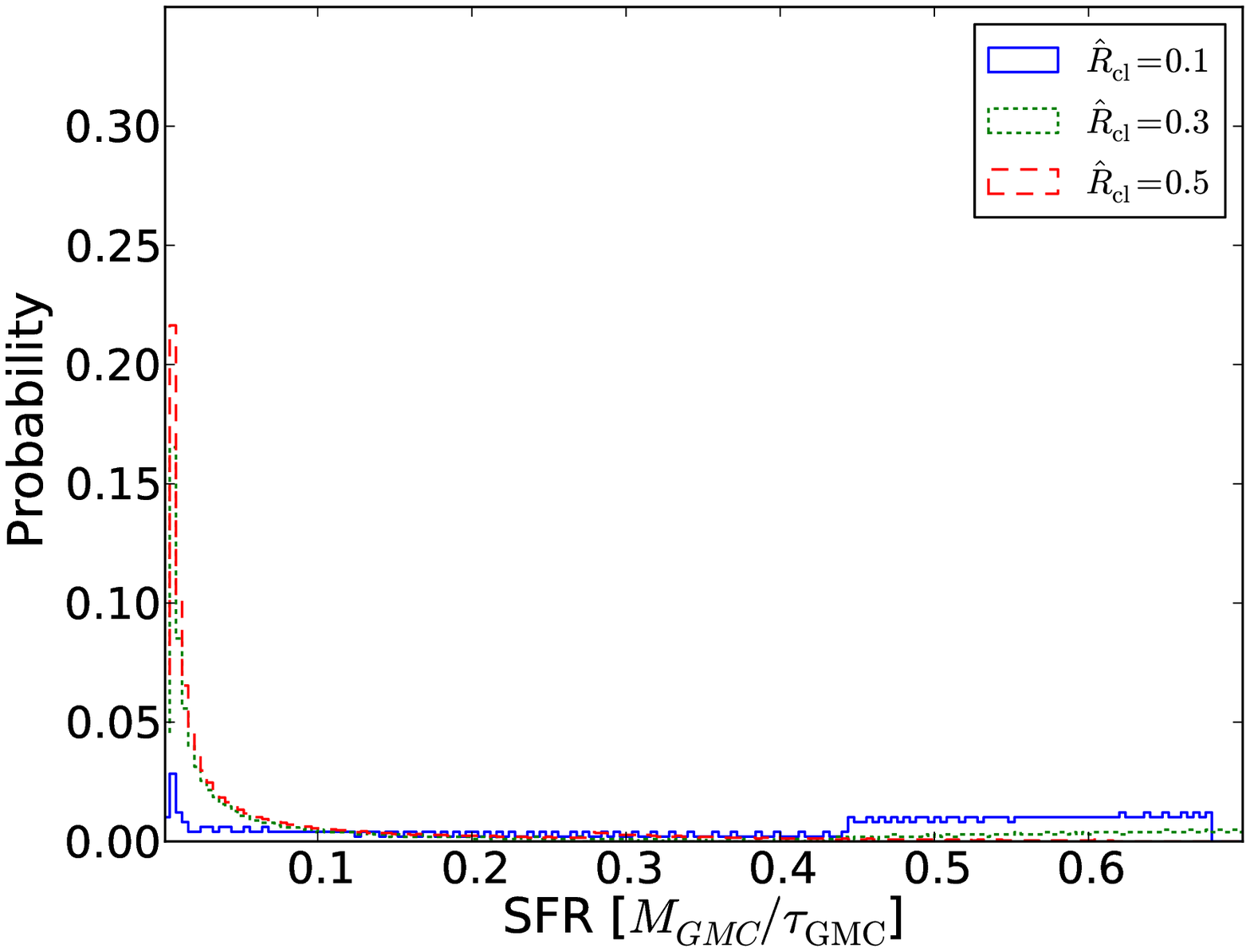}\\
\includegraphics[width=.33\textwidth]{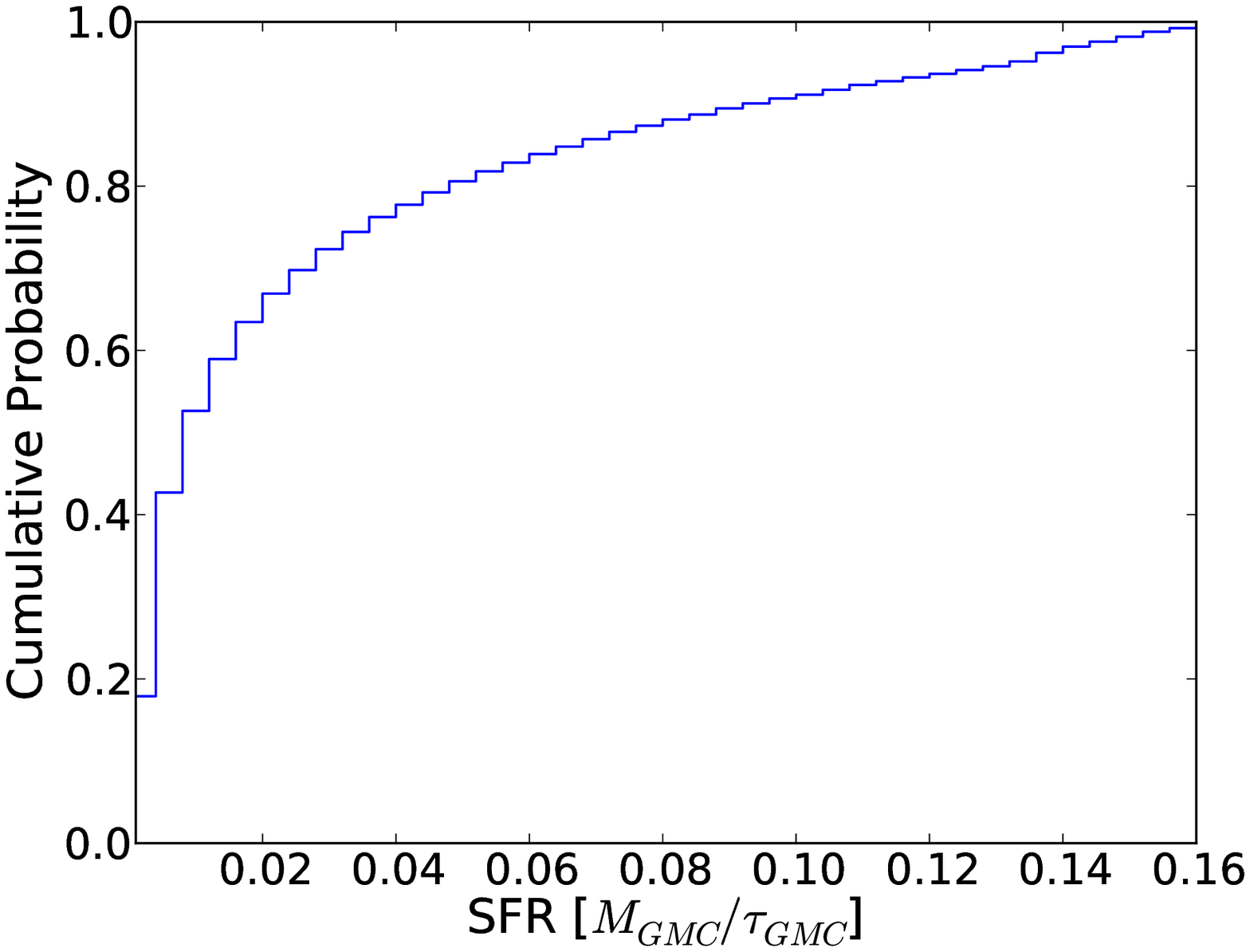}
\includegraphics[width=.33\textwidth]{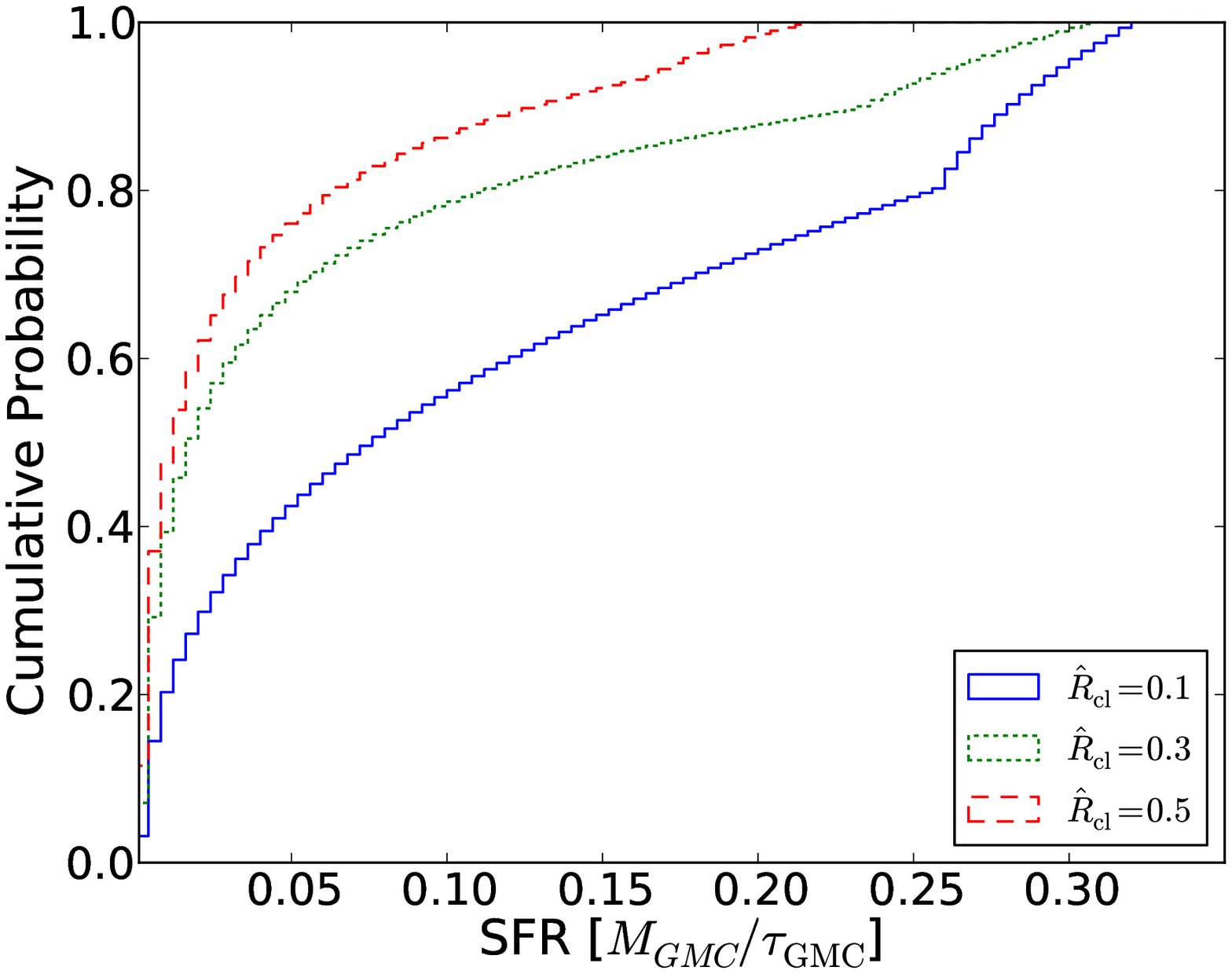}
\includegraphics[width=.33\textwidth]{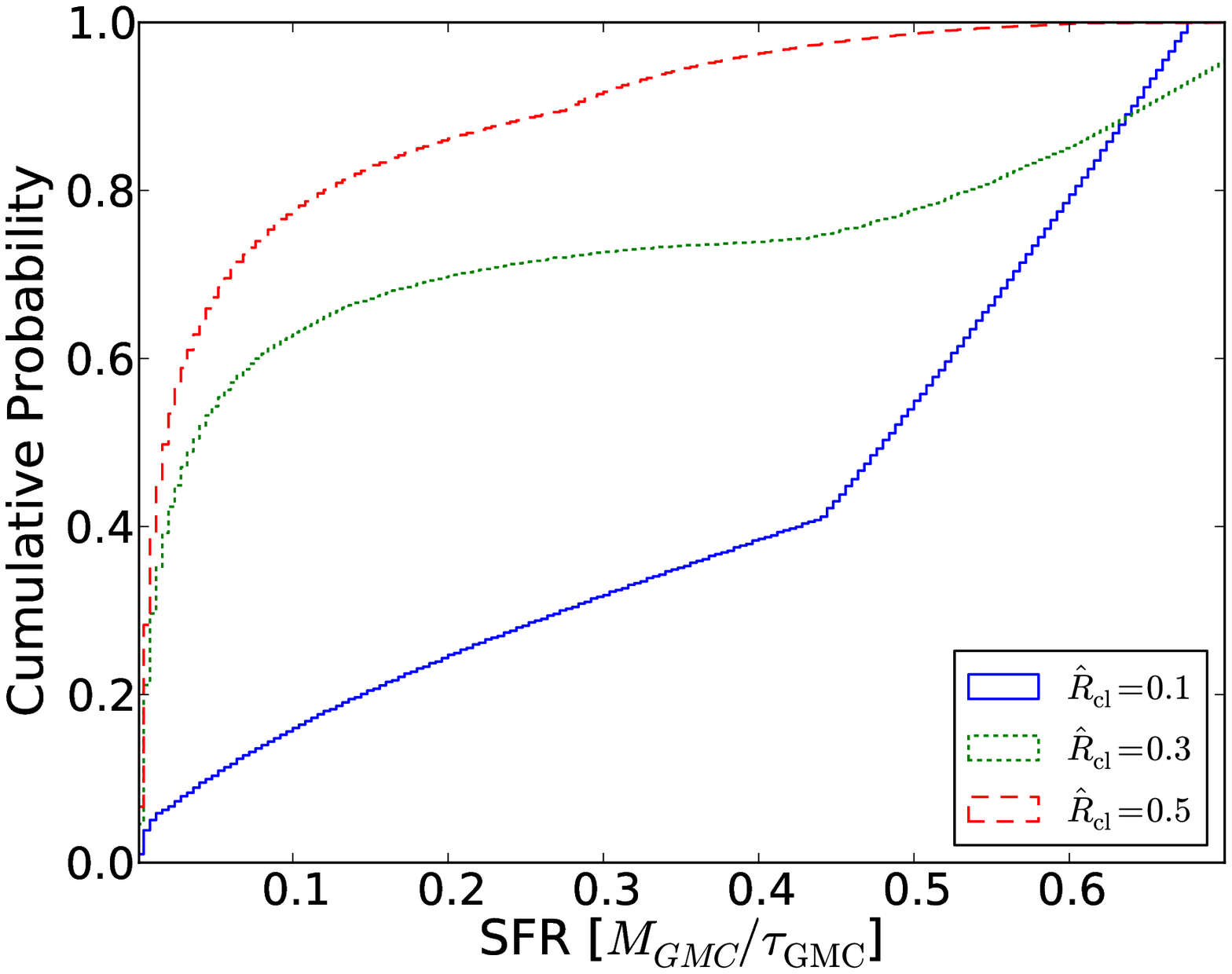}
\end{center}
\caption{Upper panels: histogram of star formation rates over the lifetime of the GMC for $\beta =
  0$ (left), $1$ (center), and $2$ (right).  Values of $\etaG\equiv \taug \dot M_*/\Mg$ below $10^{-4}$ are discarded. Lower panels: Cumulative histogram of star formation rates $\etaG$ over the lifetime of the GMC for $\beta = 0$ (left), $1$ (center), and $2$ (right). The star formation rate is low for the majority of the lifetime of the GMC, but a sizeable fraction of a GMC's life is spent at $\etaG\gtrsim 0.1$.  That fraction increases with increasing central concentration, i.e., $\beta = 1$ and $2$. In both the upper and lower panels, we show the cases of $\xclumphat=0.1$ (blue solid line), $0.3$ (green dotted line) and $0.5$ (red dashed line) in the middle ($\beta = 1$) and right ($\beta = 2$) panels.}
\label{fig:histograms} 
\end{figure*}

In the upper panels of Figure \ref{fig:histograms}, we show the histogram of star formation efficiencies per free-fall time $\etaG$  for  $\etaG> 10^{-4}$. For $\beta \neq 0$ (middle and right panels), we show the cases for $\xclumphat=0.1$ (blue solid line), $0.3$ (green dotted line) and $0.5$ (red dashed line). These panels shows that a GMC spends most of its life at a very low $\etaG\lesssim0.02$, but a significant portion of its life is spent at larger $\etaG\approx0.1$.  This is seen most clearly in the cumulative probability distribution in the lower panels of Figure \ref{fig:histograms}.  For instance, in the $\beta = 0$ case, half of the lifetime of a cloud is spent at $\etaG\lesssim 0.015$.  However, $10\%$ of the lifetime of the cloud is spent at $\etaG > 0.1$, with a maximum  $\etaG=0.16$.  

For more centrally concentrated GMCs, the fraction of GMC lifetime spent at large $\etaG$ become more pronounced.  Indeed for, $\beta = 1$ and $r = 0.1$, the GMC spends half its life with  $\etaG \gtrsim 0.08$, and $\sim10\%$ of its life is spent with $\etaG\approx 30\%$.  

The wide range in $\etaG$ of the Bondi accretion model is consistent with the observed range in $\etaG=0.002$ to $0.2$ in Galactic GMCs \citep{2011ApJ...729..133M}. Moreover, \citet{2011ApJ...729..133M} found that GMCs with $\etaG \approx 0.2$ are in the act of being disrupted.  This is consistent with the notion that stellar feedback at high $\etaG$ caps both the maximum star formation rate and the lifetime of GMCs.  

We note that these observations are less consistent with models where global turbulent statistics set the SFR (Padoan 1995; Krumholz \& McKee 2005); in these models $\eff =\eta =\etaG\approx0.017$.  Namely, for a turbulence-limited SFR, the values of $\eta$ can be expected to be roughly a factor of $3$ above and below the average value of $\eta \approx 0.02$.  However, the measure values of $\etaG$ in our galaxy range up to  $\approx 10\eff$ or higher.  It is possible that this is merely a random conspiracy of turbulent statistics, but we would argue that accretion onto clumps gives a clearer picture of how such high SFRs are achieved.  

\subsection{The effect of accretion on the mass spectrum of clumps}

The mass spectrum of both clumps and star clusters is generally fit by a powerlaw distribution,
\be \label{eq: spectrum}
{dN_{cl}\over d\ln m}=A\left({m_0\over m}\right)^{\alpha-1},
\ee 
with $\alpha\approx 1.5-2$, e.g., \citet{1989ApJ...337..761K}. If, as we argue, clumps undergo a period of rapid growth, and that more massive clumps grow more rapidly, will an initial power law distribution result in a final distribution of clumps (or star clusters) that is also a power law?

Equation (\ref{eqn:final}) shows how to map an initial clump mass into a final clump mass, given the age of the clump when accretion stops. We can map an initial clump mass function into a final clump mass function (when the host GMC is disrupted) by assuming a distribution of clump formation times. We work backwards from the time when the largest clump (or the combined effects of all the clumps) disrupts the GMC, when the largest clump has a given age. We then assume a Gaussian distribution of clump formation times with a mean equal to that of the largest cluster. To be definite, we assume that the largest cluster has a final mass of $\sim5\%$ of that of the host GMC when the latter is disrupted. We also fix the number of clusters to be in the range 20-50, with total initial mass of $0.001\Mg$. We consider cases with $1.5<\alpha<2$. 

\begin{figure}
\begin{center}
\includegraphics[width=.5\textwidth]{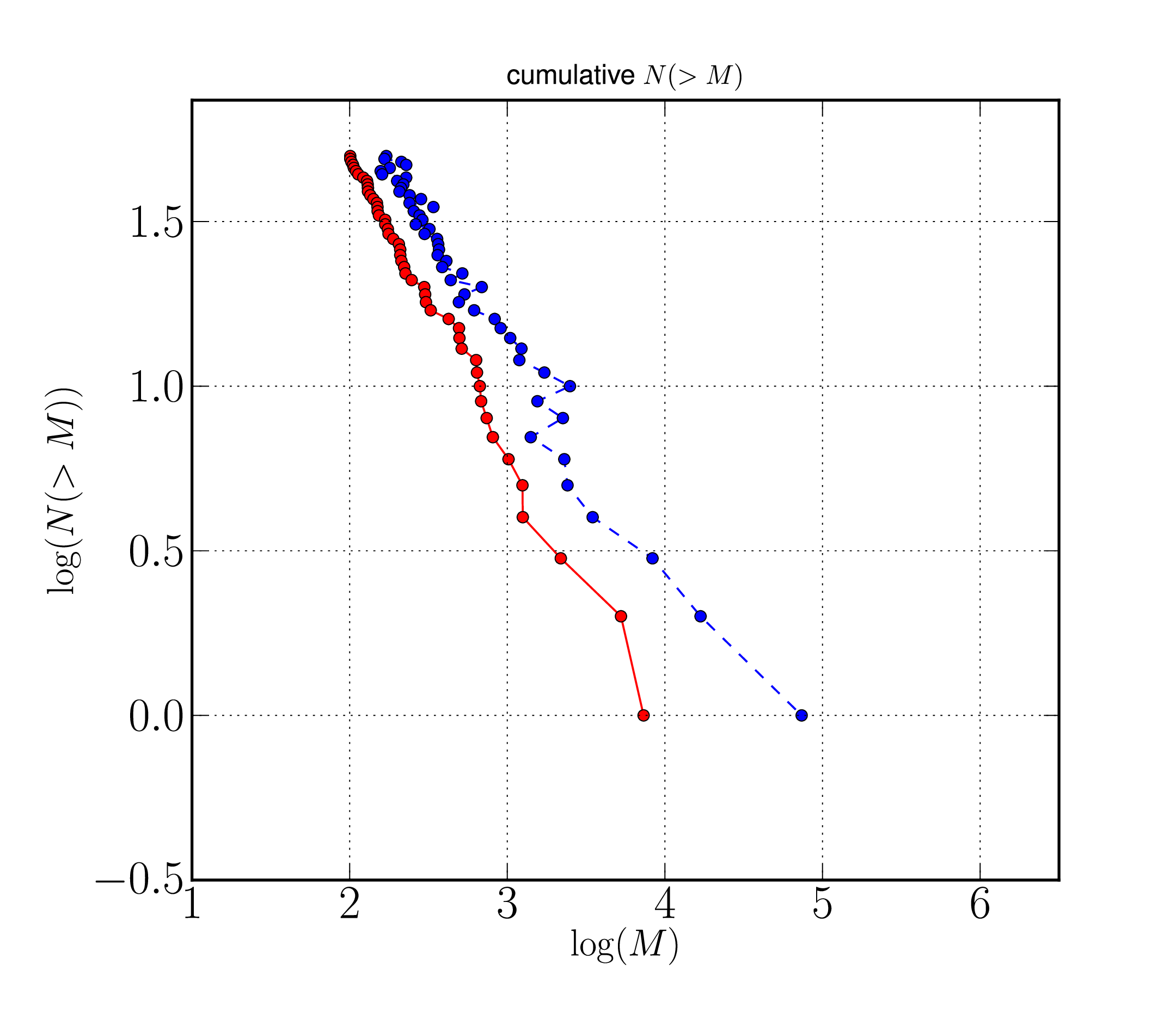}
\end{center}
\caption{The cumulative clump mass distribution function $N(>m)$. The red solid points, connected by a solid line, show the initial clump mass distribution function, while the blue points (connected by a dashed line) show the clump mass distribution function at the point when the host GMC is disrupted. The initial slope is $\alpha = 1.9$, while the final slope is $\alpha = 1.7$. The initial total clump mass is $0.03\Mg$, the final total clump mass is $0.13\Mg$. 
}
\label{fig:clump mass function}
\end{figure}

Figure \ref{fig:clump mass function} shows a typical result. The change in the power law slope $\alpha$ is $\Delta\alpha\sim 0.3$, e.g., if $\alpha=-2.0$ for the initial distribution of clump masses, the final least squares fit (when the total stellar mass is $\sim0.1\Mg$ ) is $\alpha=1.7$. More generally, we find $\Delta\alpha\approx 0.3$ for a range in the inital slopes $1.5<\alpha<2.0$.

\section{Discussion}\label{sec: discussion}

\subsection{Bondi Clumps in Numerical Simulations}
Recent simulations \citep{2009ApJ...707.1023V,2010ApJ...715.1302V}  are consistent with our picture of star cluster formation.  These authors studied the fragmentation of supersonically turbulent molecular gas into star-forming clumps (clouds in their nomenclature) and the subsequent evolution of these clumps.  In their simulation, star-forming clumps occur when transonic converging flows in diffuse warm gas cause that gas to be rapidly transformed to cold dense material.  This cold gas, which is in an initial pancake-like structure, collapses into filaments, and then to clumps. Their major result is that these star-forming clumps continue to accrete material from the background GMC at a high rate through the filaments.

Accretion via filaments is also seen in smaller scale simulations \citep{2006MNRAS.373.1091B}. In these simulations the collapse of a Bonner-Elbert sphere supported by supersonic turbulent motions is followed. The sheet-like and filamentary structures that arise from the collisions of supersonic eddies lead to regions of very high density which collapse into protostellar clumps (disks and protostars).  These authors found that the clumps continue to accrete material from the filaments at a very high rate, such that massive stars could form in a few thousand years.  Indeed the accretion rate from filamentary accretion was a factor of $10^3$ times larger than what would be expected from collapse of a singular isothermal sphere.  This discrepancy arises from the assumption that the accretion proceeds at the sound speed of the cold material; this sound speed is smaller than the velocity needed to support an isothermal sphere of the type \citep{2006MNRAS.373.1091B} simulated, since their clouds were supported by supersonic motions.

In both sets of  simulations just described, clumps are produced in regions where transonic converging flows occur. Even in the absence of gravity from the newly formed clump, these convergent large scale motions will enhance the accretion rate of the clump relative to the rate that would be expected from a similar mass clump placed at random in the simulation. We argue, however, that as the clump grows, the expanding gravitational reach of the clump will direct larger and larger amounts of gas to collapse onto the sheets and filaments, and thence onto the clump, yielding the accretion rates we have calculated. Testing this process by employing both gravitating and non-gravitating sink particles in large scale turbulence simulations is a subject of our ongoing work.





\subsection{The Effect of Vorticity}\label{sec:vorticity}
So far we have ignored the possible role of angular momentum or vorticity in limiting the rate of accretion onto clumps. If the accreting gas carries a substantial amount of angular momentum, it may become rotationally supported at radii larger than the $\sim\pc$ scale at which we assume star forming clumps are born. If this were to happen, i.e., if the 
circularization radius $r_{\rm circ}$ of the accreting gas is larger than $\rcl$, the rate of star formation would be suppressed relative to the rates we have calculated. 

We now argue that vorticity does not play a role in the formation of star clusters inside GMCs in local galaxies. 

Observational estimates of the angular momentum of GMCs start from measurements of the velocity gradient $\nabla v$ across the cloud, e.g., \citet{2003ApJ...599..258R}, and assume $\Omega \equiv|\nabla {\bf v}|$. As these authors note, strictly speaking this is an upper limit to $\Omega$, since $|\nabla {\bf v}|$ contains contributions from non-rotational motions. The measurements suggest that $\Omega(r)\approx const.$, i.e. the vorticity ($\nabla\times v$) is constant, or equivalently, that specific angular momentum $j(r)$ of a parcel of gas at radius $r$ scales as $j(r)=j(\Rg)(r/\Rg)^2$.

We characterize the rotation of the GMC by the parameter
\be  
\gamma \equiv {\Rg|\nabla {\bf v}|\over \vg}.
\ee  
%


Then
\be  
j(r) = \gamma \Rg\vg\left({r\over\Rg}\right)^2.
\ee  

We assume that $j$ does not change during the core accretion process, from its initial radius (the Bondi-Hoyle radius $\rb$) to the circularization radius $r_{\rm circ}$, so that the circularization radius of a parcel of gas initially at radius $r$ accreting onto a clump of mass $\mcl$ is given by $j(r)=\sqrt{G\mcl r_{\rm circ}}$. We require that $r_{\rm circ}<\rcl$, the cluster radius, leading to the critical value of the parameter $\gamma$ for a GMC with $\rho(r)\sim r^{-\beta}$ (c.f. Equation (\ref{eq:density})) of
\be  
\gamma_{\rm crit}=\mu^{-(1+\beta)/2(3-\beta)}\sqrt{\rcl\over\Rg}.
\ee  

If a GMC has $\gamma\le\gamma_{\rm crit}$, then gas will accrete onto a cluster of radius $\rcl$ before the gas becomes rotationally supported, and the accretion can be described by Equation (\ref{eqn:mdot Bondi}).
Observations of GMCs in M33 find typical values $\Rg\approx23\pc$ \citep{2003ApJ...599..258R}; scaling to $\mu=0.1$ and $\beta =1$, we find $\gamma_{\rm crit}=0.66(0.1/\mu)^{1/2}(\rcl/1\pc)^{1/2}$. 
Typical values for the specific angular momentum of GMCs in M33 are $j\approx22\pc\kms$, corresponding to $\gamma\approx0.17$. Since $\gamma<\gamma_{\rm crit}$ it follows that clusters in M33 GMCs can grow to $\sim10\%$ of the host GMC mass before vorticity begins to affect the accretion process.
The situation is similar in M31, with $\Rg\approx34\pc$ \citep{2007ApJ...654..240R}, leading to $\gamma_{\rm crit}=0.55(0.1/\mu)^{1/2}$, while the observed value is $\gamma\approx 0.28$. 
Thus, we conclude that vorticity in GMCs is unlikely to limit the rate of accretion onto self-gravitating clumps, and hence the rate of star formation.

\section{CONCLUSIONS}
We have argued that star formation in turbulent GMCs, which is observed to occur in massive dense clumps, is in fact controlled by the properties of accretion onto those  clumps. We start by assuming that the turbulence generates shocks, which form filaments and subsequently massive dense clumps. These turbulently generated clumps act as the initial conditions for the subsequent gravitationally dominated accretion phase. We then used the observed properties of the internal gas motions in GMCs (Larson's law, $v_T(l)\sim l^{1/2}$) to modify the Bondi argument for the accretion rate of a massive dense clump embedded in that gas; for a clump located slightly away from the center of the GMC with a non-uniform density distribution,  or for a uniform density GMC, that rate is given by Equation (\ref{eq: general mdot}). Left to their own devices, these clumps can accrete all the gas in the GMC in only a few GMC free-fall times. 

We then showed that the star formation rate in an individual dense clump tends to track the mass accretion rate. The gas fraction in the cluster approaches a psuedo-fixed point, given by Equation (\ref{eq: fixed point}). 

The model predicts that the distribution of star formation rates in GMCs should peak at low values $\etaG\lesssim 0.01$, but that there is a substantial tail to high star formation rates, with $\etaG\gtrsim 0.1$, i.e., ten percent of the GMC is converted to stars in a single GMC free-fall time. This is in contrast to theories of turbulence regulated star formation, but agrees with recent measurements of $\etaG$ on small spatial scales, e.g., \citet{2010ApJ...722.1699S}, which show a broad distribution. A second prediction of the model is that the star cluster mass function is flatter (has a smaller index $\alpha$ in Equation (\ref{eq: spectrum})) than the mass function of clumps, though it is still relatively close to the initial clump mass function.  These aspects of our model, i.e., the broad distribution in star formation rates, the association of the largest star formation rates with the largest star clusters and GMC disruption, and a star cluster mass function with a power law slope of around $\alpha \sim 1.5-2$, are in broad agreement with observations.

\acknowledgments
N.M. and P.C. are supported in part by NSERC of Canada. N.M.~is supported in part by the Canada Research Chair program. This research has made use of NASA's Astrophysics Data System.
%




\bibliography{Bondi}{}

\end{document}